\documentclass[twocolumn,nofootinbib]{revtex4}
\usepackage{graphicx}
\usepackage{bm}
\usepackage{amsmath}
\usepackage[dvipsnames]{color}

\newcommand{\kslash}{k\hspace{-0.075in}\slash\hspace{0in}}
\newcommand{\pslash}{p\hspace{-0.06in}\slash\hspace{-0.015in}}

\begin{document}
\title{Unitarity and Bounds on the Scale of Fermion Mass Generation}
\author{R. Sekhar Chivukula}
\email[email: ]{sekhar@msu.edu}
\author{Neil D. Christensen}
\email[email: ]{neil@pa.msu.edu}
\author{Baradhwaj Coleppa}
\email[email: ]{baradhwaj@pa.msu.edu}
\author{Elizabeth H. Simmons}
\email[email: ]{esimmons@msu.edu}
\affiliation{Department of Physics and Astronomy\\ Michigan State University\\ East Lansing, MI 48824}
\date{\today}

\begin{abstract}
The scale of fermion mass generation can,
as shown by Appelquist and Chanowitz, be bounded {\it from above} by relating it
to the scale of unitarity violation in the helicity nonconserving amplitude for fermion-anti-fermion
pairs to scatter into pairs of longitudinally polarized electroweak gauge bosons.
In this paper, we examine the process $t\bar{t} \to W_L^+ W_L^-$  in a family of phenomenologically-viable deconstructed Higgsless models and we show that scale of unitarity violation depends on the mass of the additional vector-like fermion
states that occur in these theories (the states that are the
deconstructed analogs of Kaluza-Klein partners of the ordinary
fermions in a five-dimensional theory). For sufficiently light vector
fermions, and for a deconstructed theory with sufficiently many lattice
sites (that is, sufficiently close to the continuum limit),  the Appelquist-Chanowitz bound can be substantially weakened.  More precisely, we find
 that, as one varies the mass of the vector-like fermion for fixed top-quark and gauge-boson masses, the bound on the scale of top-quark mass generation interpolates smoothly between the Appelquist-Chanowitz
bound and one that
can, potentially, be much higher.  In these theories, therefore, the bound on the scale of fermion
mass generation is independent of the bound  on the scale of gauge-boson
mass generation. While our analysis focuses on deconstructed Higgsless models,
any theory in which top-quark mass generation
proceeds via the mixing of chiral and vector fermions will give similar
results.
\end{abstract}

\maketitle

\section{Introduction}

Although the mechanism of electroweak symmetry breaking remains a mystery, it is
clear that this mechanism must give mass to two very different classes of 
particles: the electroweak gauge bosons and the fermions. In the standard
model, the scalar Higgs \cite{Higgs:1964ia} doublet couples directly to both classes of 
particles \cite{Weinberg:1967tq,Salam:1968rm}.
Moreover, the gauge and Yukawa couplings through which the Higgs interacts, respectively, with
gauge bosons and fermions are proportional to the masses generated for those states when the scalar
doublet acquires a vacuum expectation value.  Nonetheless, in considering physics beyond the
standard model, the possibility remains that the gauge boson and fermion masses are generated through
different mechanisms. In particular, it is possible that electroweak symmetry breaking
is transmitted to the fermions via some intermediary physics specifically associated
with  fermion mass generation.

Appelquist and Chanowitz  \cite{Appelquist:1987cf}
have shown\footnote{See also \protect\cite{Marciano:1989ns,Golden:1994pj}.}
that the tree-level,  spin-0 scattering amplitude for fermion-anti-fermion pairs 
to scatter into longitudinally-polarized electroweak gauge bosons grows
linearly with energy below the scale of the physics responsible for transmitting
electroweak symmetry breaking to the fermions. As the amplitude must be unitary,
one can derive an {\it upper} bound on the scale of fermion mass generation by
finding the energy at which the amplitude would grow to be of order 1/2 .
The rate of energy growth is proportional to the mass of the fermions involved.
The most stringent bound, therefore, arises from top-quark annihilation,
and the bound on
the scale of top-quark mass generation is found to be of order a few TeV.\footnote{For
light fermions, the scattering of fermions into many gauge-bosons yields a stronger
result than the Appelquist-Chanowitz bound \protect\cite{Maltoni:2001dc,Dicus:2004rg}. For
the top-quark, however, two-body final states yield the strongest bound.} 

As emphasized by Golden \cite{Golden:1994pj}, the interpretation of the Appelquist-Chanowitz
(AC) bound on the scale of top-quark mass generation can be problematic: longitudinal electroweak gauge-boson elastic scattering
itself grows quadratically with energy \cite{LlewellynSmith:1973ey,Dicus:1992vj,Cornwall:1974km,Lee:1977eg,Veltman:1976rt} below the scale of the physics responsible for electroweak gauge-boson
mass generation. As the scale of the physics responsible for electroweak symmetry
breaking is also bounded by of order a TeV, it can be difficult to be sure that the violation of unitarity in fermion annihilation
is truly independent of the violation of unitarity in the gauge-boson sector. The standard model
illustrates this difficulty, as in that case the Higgs boson is responsible for restoring
unitarity in {\it both} the fermion annihilation and gauge-boson scattering processes.

In this paper, we discuss unitarity violation and the resulting bounds on the scale of top-quark mass generation in the context of deconstructed Higgsless models. Higgsless models
\cite{Csaki:2003dt} achieve electroweak
symmetry breaking without introducing a fundamental scalar Higgs
boson \cite{Higgs:1964ia}, and the unitarity of longitudinally-polarized
$W$ and $Z$ boson scattering  \cite{LlewellynSmith:1973ey,Dicus:1992vj,Cornwall:1974km,Lee:1977eg,Veltman:1976rt} is preserved by the exchange of extra vector
bosons \cite{SekharChivukula:2001hz,Chivukula:2002ej,Chivukula:2003kq,He:2004zr}.
Inspired by TeV-scale \cite{Antoniadis:1990ew} compactified five-dimensional
gauge theories  \cite{Agashe:2003zs,Csaki:2003zu,Burdman:2003ya,Cacciapaglia:2004jz}, these models provide effectively unitary descriptions of the electroweak sector beyond 1 TeV. 
Deconstruction \cite{Arkani-Hamed:2001ca,Hill:2000mu} is a technique
to build a four-dimensional gauge theory, with an appropriate gauge-symmetry breaking pattern,
which approximates the properties of a five-dimensional theory. Deconstructed Higgsless models \cite{Foadi:2003xa,Hirn:2004ze,Casalbuoni:2004id,Chivukula:2004pk,Perelstein:2004sc,Georgi:2004iy,SekharChivukula:2004mu} have been used as tools to compute the general properties of Higgsless theories, and to illustrate the phenomenological properties of this class of models. 

The simplest deconstructed Higgsless model \cite{SekharChivukula:2006cg,Casalbuoni:1985kq} incorporates only three sites on the deconstructed lattice, and the only additional vector states (other than the
usual electroweak gauge bosons) are a triplet of vector bosons.
While simple, the three site model  is
sufficiently rich to describe the physics associated with fermion mass generation,
as well as the  fermion delocalization \cite{Anichini:1994xx,Cacciapaglia:2004rb,Cacciapaglia:2005pa,Foadi:2004ps,Foadi:2005hz,Chivukula:2005bn,Casalbuoni:2005rs,SekharChivukula:2005xm}
required in order to accord with precision electroweak tests
\cite{Peskin:1992sw,Altarelli:1990zd,Altarelli:1991fk,Barbieri:2004qk,Chivukula:2004af}.
It is straightforward to generalize this model to an arbitrary number of
sites \cite{Coleppa:2006fu}.  In the continuum limit (the limit in which the number of sites goes
to infinity), this model reproduces the five-dimensional model introduced in \cite{Foadi:2005hz}.

A fermion field in a general compactified five-dimensional theory gives rise to
a tower of Kaluza-Klein (KK) modes, the lightest of which can (under chiral  boundary
conditions) be massless in the absence of electroweak symmetry breaking.  The lightest states
can therefore be identified with the ordinary fermions. The massive
Kaluza-Klein fermion modes are, however, massive Dirac fermions
from the four-dimensional point of view. Correspondingly, 
the fermions in a deconstructed Higgsless model include both chiral and vector-like
electroweak states \cite{SekharChivukula:2006cg,Coleppa:2006fu}, and 
generation of the masses of the ordinary fermions in these models involves the mixing of the chiral
and vector states \cite{Cacciapaglia:2005pa,Foadi:2004ps}. As we will
demonstrate, the scale of top-quark mass 
generation in these models depends on the masses of the vector-like fermions
(the ``KK" modes), as well as on the number of sites in the deconstructed lattice.

What is particularly interesting about deconstructed Higgsless
models, in this context, is that one {\it can} distinguish between the unitarity-derived bounds on the scales of 
gauge-boson and top-quark mass generation. We will demonstrate that, for an appropriate number
 of deconstructed lattice sites, spin-0 top-quark annihilation to longitudinally-polarized
gauge-bosons remains unitary at tree-level up to energies much higher than the naive
AC bound if the vector-like fermions are light.
However the AC bound is reproduced as the mass of the vector-like fermion is
increased. Therefore, for fixed top-quark and gauge-boson masses, the bound on the scale of
fermion mass generation interpolates smoothly between the AC bound and one that
can, potentially, be much higher as the mass of the vector-like
fermion varies.  The unitarity bounds on elastic scattering of longitudinal electroweak gauge bosons in Higgsless models \cite{SekharChivukula:2006we}, however, depend only on the masses
of the gauge-boson KK modes.
 In this sense, the bound on the scale of fermion
mass generation is {\it independent}  of the bound  on the scale of gauge-boson
mass generation.

While our discussion is restricted to deconstructed Higgsless models, many models 
of dynamical electroweak symmetry breaking incorporate
the mixing of chiral and vector fermions to accommodate top-quark mass generation.
Examples include the top-quark seesaw model \cite{Dobrescu:1997nm,Chivukula:1998wd,He:2001fz},
and models in which the top mixes with composite fermions arising from a dynamical electroweak
symmetry breaking sector \cite{Suzuki:1991kh,Lebed:1991qv,Kaplan:1991dc}.
Indeed, the fermion delocalization required to construct a realistic Higgsless model
is naturally interpreted, in the context of AdS/CFT duality  \cite{Maldacena:1998re,Gubser:1998bc,Witten:1998qj,Aharony:1999ti}, as mixing between fundamental and composite
fermions \cite{Csaki:2005vy}. As chiral-vector fermion mixing is the basic feature
required for our results, we expect similar effects  in these other models.

In the next section, to set notation and make contact with the literature, 
we reproduce \cite{Golden:1994pj} the Appelquist-Chanowitz bound in the electroweak
chiral Lagrangian \cite{Appelquist:1980ae,Appelquist:1980vg,Longhitano:1980iz,Longhitano:1980tm,Appelquist:1993ka}  --- which may be interpreted as a ``two-site" 
Higgsless model. In section three, we introduce the $n(+2)$ site Higgsless models that we will use
for our calculations. Section four contains our calculations and primary results. The
last section summarizes our findings.

\section{\label{The Appelquist-Chanowitz Bound}The Appelquist-Chanowitz Bound}

\begin{figure}
  \begin{center}
    \includegraphics{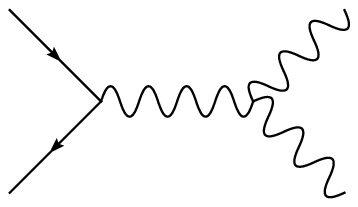}
    \begin{picture}(0,0)(0,0)
      \put(-120,60){$t_+$}
      \put(-120,-5){$\bar{t}_+$}
      \put(0,60){$W_L^+$}
      \put(0,-5){$W_L^-$}
      \put(-70,40){$\gamma,Z$}
    \end{picture}
    \vspace{0.3in}\\
    \includegraphics[]{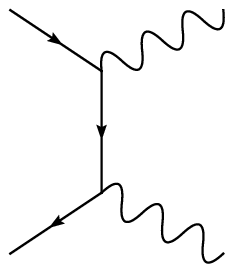}
    \begin{picture}(0,0)(0,0)
      \put(-80,75){$t_+$}
      \put(-80,0){$\bar{t}_+$}
      \put(0,75){$W_L^+$}
      \put(0,0){$W_L^-$}
      \put(-35,40){$b$}
    \end{picture}
    \caption{\label{ttbar -> WW}The diagrams that contribute to the process $t_+\bar{t}_+\rightarrow W_L^+W_L^-$ in the Higgsless SM.  There are analogous diagrams for the process $t_-\bar{t}_-\rightarrow W_L^+W_L^-$.  Each diagram has an amplitude that grows linearly with $\sqrt{s}$ for all energies.  However, most (but not all) of this linear $\sqrt{s}$ growth cancels when the diagrams are summed.  The remaining piece that grows linearly with $\sqrt{s}$ comes from the t channel diagram, and it eventually surpasses the unitarity bound.  In the SM, this unitarity violation is eliminated by the contribution of the Higgs in the $s$ channel.}
  \end{center}
\end{figure}

In the standard model (SM), the helicity non-conserving process 
$t_+\bar{t}_+\rightarrow W_L^+W_L^-$ receives contributions at tree level from the diagrams in Figure \ref{ttbar -> WW}.
We are interested in the behavior of the amplitude for large  center of mass energy,  $\sqrt{s}\gg M_W,m_t$.  This allows us to expand the amplitude in the small parameters $M_W^2/s$ and $m_t^2/s$.  Practically, this means that we use  the following leading order approximations.  For the longitudinal polarization of the $W$ gauge boson, we use   
\begin{equation}\label{epsilon_{W_L}}
\epsilon_{W_L}^\mu \simeq \frac{k_{W_L}^\mu}{M_W},
\end{equation}
where $k^\mu_{W_L}$ is the four-momentum of the corresponding boson.
For the spinor chain in the $s$ channel, we use
\begin{eqnarray}
\bar{v}_+\left(\kslash_1-\kslash_2\right)\left(g_LP_L+g_RP_R\right)u_+ 
&\simeq& m_t \sqrt{s}\cos\theta\left(g_L+g_R\right)\nonumber\\\label{v+u+ s}\\
\bar{v}_-\left(\kslash_1-\kslash_2\right)\left(g_LP_L+g_RP_R\right)u_- 
&\simeq& -m_t \sqrt{s} \cos\theta\left(g_L+g_R\right)~,\nonumber\\\label{v-u- s}
\end{eqnarray}
where $k_1^\mu$ and $k^\nu_2$ are the momenta of the outgoing bosons,
and for the spinor chain in the $t$ channel we find
\begin{eqnarray}
\bar{v}_+\kslash_2\left(\pslash_1-\kslash_1\right)\kslash_1g_LP_Lu_+ 
&\simeq& \frac{m_t t \sqrt{s}}{2}\left(1+\cos\theta\right)g_L\nonumber\\\label{v+u+ t}\\
\bar{v}_-\kslash_2\left(\pslash_1-\kslash_1\right)\kslash_1g_LP_Lu_-
&\simeq& -\frac{m_t t \sqrt{s}}{2}\left(1+\cos\theta\right)g_L\nonumber\\\label{v-u- t}
\end{eqnarray}
where
\begin{eqnarray}
P_L&=&\frac{1}{2}\left(1-\gamma_5\right)\\
P_R&=&\frac{1}{2}\left(1+\gamma_5\right)
\end{eqnarray}
are chirality projection operators, and $g_L$ and $g_R$ are chiral electroweak coupling constants.

Since the $t \bar{t} \to W^+ W^-$ amplitude is the same for each color and only differs by a sign for the opposite helicity, we get the largest amplitude by considering the incoming state\footnote{The state we 
consider here differs from that chosen by \protect\cite{Appelquist:1987cf}, as we include both combinations of incoming helicities. This state allows us to derive a slightly stronger bound,
{\it c.f.} Eqn (\protect\ref{eq:ouracresult}).}
\begin{eqnarray}
\left| \psi \right> = \frac{1}{\sqrt{6}}\Big(\label{in state}&
\left| \bar{t}_{1+}t_{1+}\right> +\ \left| \bar{t}_{2+}t_{2+}\right> +\ \left|\bar{t}_{3+}t_{3+}\right>&\\
&-\ \left|\bar{t}_{1-}t_{1-}\right> -\ \left| \bar{t}_{2-}t_{2-}\right> -\ \left|\bar{t}_{3-}t_{3-}\right>&
\Big)\nonumber~,
\end{eqnarray}
where the numerical subscripts (1,2, and 3)  label the three different colors.
Putting the pieces together gives the scattering amplitude
\begin{eqnarray}
& &\mathcal{M} (\psi \to W_L W_L) = \frac{\sqrt{6}\,m_t\sqrt{s}\cos\theta}{2M_W^2}\\
&&\times\left( 2 g_{tt\gamma}g_{\gamma WW} + g_{LttZ}g_{ZWW} + g_{RttZ}g_{ZWW} - g_{LtbW}^2 
\right)\nonumber\\
&&+\frac{\sqrt{6}\,m_t\sqrt{s}}{2M_W^2}g_{LtbW}^2~,\nonumber
\end{eqnarray}
for $\sqrt{s} \gg M_W,\, m_t$, where the
electroweak couplings are given by\footnote{Our expression here differs in the sign of the
term proportional to $g_{LtbW}^2$ from that given in \protect\cite{Appelquist:1987cf},
and is correct for the top-quark which is the $T_3=+1/2$ member of an electroweak
doublet. The corresponding expression in 
\protect\cite{Appelquist:1987cf}, which is from \protect\cite{Chanowitz:1978uj,Chanowitz:1978mv}, is
correct for the {\it lower} member of an electroweak doublet with $T_3=-1/2$.}
\begin{eqnarray}
g_{tt\gamma} &=& \frac{2}{3}e~, \\
g_{\gamma WW} &=& e~,  \\
g_{LttZ} &=& \frac{e}{\sin\theta_W \cos\theta_W} \left( \frac{1}{2} - \frac{2}{3}\sin^2\theta_W \right)~, \\ 
g_{RttZ} &=& \frac{e}{\sin\theta_W \cos\theta_W} \left(- \frac{2}{3}\sin^2\theta_W \right)~, \\
g_{ZWW} &=& \frac{e\cos\theta_W}{\sin\theta_W}~, \\ 
g_{LtbW} &=& \frac{e}{\sqrt{2}\sin\theta_W}~. 
\end{eqnarray}
With these couplings, we find the identity
\begin{equation}
2 g_{tt\gamma}g_{\gamma WW} + g_{LttZ}g_{ZWW} + g_{RttZ}g_{ZWW} - g_{LtbW}^2 = 0~.
\end{equation}
The remaining amplitude is, therefore,
\begin{equation}
\mathcal{M} = \frac{\sqrt{6}\,m_t\sqrt{s}}{2M_W^2}g_{LtbW}^2
\end{equation}
which grows linearly with $\sqrt{s}$ for $\sqrt{s}\gg M_W,m_t$.  We note that $g_{LtbW} = g/\sqrt{2}$ and $M_W=gv/2$, where $g$ is the weak coupling and
$v\simeq 246\,{\rm GeV}$ is the weak scale, giving \cite{Appelquist:1987cf}
\begin{equation}
\mathcal{M} = \frac{\sqrt{6}\,m_t\sqrt{s}}{v^2}~.
\label{eq:mcall}
\end{equation}

\begin{figure}
  \begin{center}
    \includegraphics[]{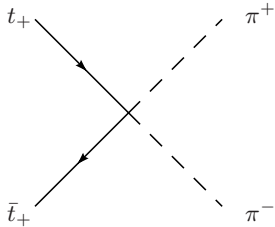}
    \begin{picture}(0,0)(0,0)
      \put(-85,70){$t_+$}
      \put(-85,-5){$\bar{t}_+$}
      \put(5,70){$\pi^+$}
      \put(5,-5){$\pi^-$}
    \end{picture}
  \end{center}
\caption{\label{AC: 2f--2phi}The diagram that contributes linear  growth in $\sqrt{s}$  to the process $t_+\bar{t}_+\rightarrow \pi^+\pi^-$ in the Higgsless SM, where we have used the equivalence theorem to 
replace the longitudinally polarized gauge-boson by the corresponding ``eaten" Goldstone Bosons.  There is an analogous diagram for the process $t_-\bar{t}_-\rightarrow \pi^+\pi^-$.}
\end{figure}

\begin{figure}
  \begin{center}
    \includegraphics[]{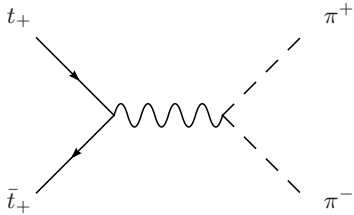}
    \begin{picture}(0,0)(0,0)
      \put(-115,65){$t_+$}
      \put(-115,-5){$\bar{t}_+$}
      \put(5,65){$\pi^+$}
      \put(5,-5){$\pi^-$}
    \end{picture}
  \end{center}
\caption{\label{AC: 2f-G-2pi}
This diagram, corresponding to $s$-channel $Z$-boson exchange in the equivalence-theorem limit, {\it does not }contribute to the $J=0$ partial wave scattering amplitude for  the process $t_+\bar{t}_+\rightarrow \pi^+\pi^-$ in the Higgsless SM.  }
\end{figure}

We can check using the equivalence theorem
\cite{Cornwall:1974km,Vayonakis:1976vz}, where one replaces
the longitudinal gauge-bosons by the corresponding ``eaten" Nambu-Goldstone Bosons.  In this
limit, the only diagram that contributes to the $J=0$ amplitude is shown in Figure \ref{AC: 2f--2phi}.
The leading order approximations
\begin{equation}
\bar{v}_+u_+  \simeq \sqrt{s} \qquad\qquad \bar{v}_-u_-  \simeq -\sqrt{s}
\end{equation}
combined with the four point coupling 
\begin{equation}
g_{tt\pi^+\pi^-} = \frac{m_t}{v^2}
\end{equation}
yield the same amplitude as in Eqn. (\ref{eq:mcall})
\begin{equation}
\mathcal{M} = \frac{\sqrt{6}\,m_t\sqrt{s}}{v^2}~.
\end{equation}
Note that the potential $s$-channel contribution, illustrated in Figure
\ref{AC: 2f-G-2pi}, does {\it not} contribute in the $J=0$ channel.

The $J=0$ partial wave is extracted from Eqn. (\ref{eq:mcall}) as 
\begin{equation}
a_0 = \frac{1}{32\pi}\int_{-1}^1 d\cos\theta\ \mathcal{M} = \frac{m_t\sqrt{6s}}{16\pi v^2}
\end{equation}
To satisfy partial wave unitarity, this tree-level amplitude must
be less than $1/2$, the maximum value for the real part of any amplitude
lying in the Argand circle.  This produces the bound 
\begin{equation}
\sqrt{s} \lesssim \frac{8\pi v^2}{m_t\sqrt{6}} \approx 3.5\,{\rm TeV}~.
\label{eq:ouracresult}
\end{equation}
Our result differs numerically from that given in \cite{Appelquist:1987cf},
as we include both helicity channels in Eq. \ref{in state}, and bound the amplitude by
1/2 rather than 1.\footnote{One may obtain a slightly stronger upper bound by considering 
an isosinglet, spin-0, final state  ($I=J=0$) of gauge-bosons \protect\cite{Marciano:1989ns}.
This amounts to a reduction in the value of the upper bound in 
Eqn. (\protect\ref{eq:ouracresult}) by a factor of $\sqrt{2/3} \approx 0.8$.}

\section{The $n(+2)$ Site Deconstructed Higgsless Model}

We will be studying the Higgsless model introduced in \cite{Coleppa:2006fu}, denoted
the $n(+2)$ site model.  As we will discuss in subsection \ref{Gauge Boson Sector}, the gauge sector is an $SU(2)^{n+1}\times U(1)$ extended electroweak group; the label $n$ thus denotes how many extra $SU(2)$ groups the model contains relative to the Standard Model.  The electroweak chiral
lagrangian \cite{Appelquist:1980ae,Appelquist:1980vg,Longhitano:1980iz,Longhitano:1980tm,Appelquist:1993ka} can be obtained by setting $n=0$ while the Higgsless Three Site Model \cite{SekharChivukula:2006cg}, which has one extra $SU(2)$ group, can be obtained by setting $n=1$.  This model may be schematically represented by a  ``Moose'' diagram \cite{Georgi:1985hf}
as shown in Figure \ref{n-moose}.  After discussing the gauge sector, we examine the fermion sector (subsection  \ref{Fermion Sector}), the ``eaten Nambu-Goldstone bosons" (subsection  \ref{Goldstone Boson Sector}) and then the couplings that are relevant to our calculation of $t\bar{t} \to  W^+ W^-$.  

\subsection{\label{Gauge Boson Sector}Gauge Boson Sector}
\begin{figure}
\begin{center}
\includegraphics[scale=1.15]{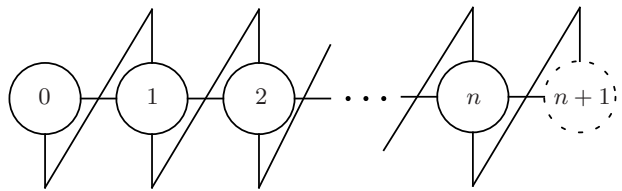}
\begin{picture}(0,0)
\put(-222,32){$0$}
\put(-181,32){$1$}
\put(-140,32){$2$}
\put(-60,32){$n$}
\put(-27,32){$n+1$}
\end{picture}
\end{center}
\caption{Moose \protect\cite{Georgi:1985hf} diagram of the $n(+2)$
site model. Each solid (dashed) circle represents an $SU(2)$ ($U(1)$)
gauge group. Each horizontal line is a non-linear sigma model. Vertical lines
are fermions, and diagonal lines represent Yukawa couplings. \label{n-moose}}
\end{figure}
The gauge group of the $n(+2)$ site model, as illustrated in Figure \ref{n-moose}, is
\begin{equation}
G = SU(2)_0 \times \prod_{j=1}^n SU(2)_j \times U(1)_{n+1}
\end{equation}
where $SU(2)_0$ is represented by the leftmost circle  and has coupling $g$; the gauge groups $SU(2)_j$ are represented consecutively by the internal circles and have a common coupling\footnote{Common couplings for the ``internal" $SU(2)$ groups corresponds to a continuum model
with spatially independent gauge-coupling \protect\cite{Foadi:2005hz}. Qualitatively, our
results do not depend on this assumption and should apply in any case
in which the mass of the $W$-boson is much less than that of the first  gauge-boson $KK$
mode.} $\tilde{g}$; 
and $U(1)_{n+1}$ is represented by the dashed circle at the far right and has coupling $g'$.  The coupling $\tilde{g}$ is taken to be much larger than $g$,  so we expand in the small quantity
\begin{equation}\label{x def}
x = \frac{g}{\tilde{g}}.
\end{equation}
We also find it convenient to define the parameters
\begin{equation}
t = \frac{g'}{g} = \frac{s}{c}
\end{equation}
where $s^2+c^2=1$. In the continuum limit, $n \to \infty$, this model reduces to the one 
described in \cite{Foadi:2005hz}.

The horizontal bars in Figure \ref{n-moose} represent nonlinear sigma models $\Sigma_j$ which 
 break the gauge symmetry down to electromagnetism
\begin{equation}
G \longrightarrow U(1)_{EM}
\end{equation}
giving mass to the other $3(n+1)$ gauge bosons.  To leading order, the effective Lagrangian
for these fields is
\begin{equation}
\mathcal{L}_{D\Sigma} = \frac{f^2}{4}\mbox{Tr}\left[\sum_j \left(D_\mu\Sigma_j\right)^\dagger D^\mu\Sigma_j\right]\label{L_D Sigma}
\end{equation}
where
\begin{equation}
D_\mu\Sigma_j = \partial_\mu\Sigma_j +ig_jW_{j,\mu}\Sigma_j - ig_{j+1}\Sigma_jW_{j+1,\mu}
\label{D_mu Sigma_j}
\end{equation}
with $g_0=g$, $g_j=\tilde{g}$ and $g_{n+1}=g'$.  
The nonlinear sigma model fields may be written
\begin{equation}
\Sigma_j = e^{i2\pi_j/f}~,
\end{equation}
in terms of  the Goldstone bosons ($\pi_j$) which become the longitudinal components of the massive gauge bosons.  The $\pi_j$ and $W_j$ are written in matrix form and are
\begin{eqnarray}
\pi_j &=& \left(\begin{array}{cc}
\frac{1}{2}\pi_j^0&\frac{1}{\sqrt{2}}\pi_j^+\\\frac{1}{\sqrt{2}}\pi_j^-&-\frac{1}{2}\pi_j^0
\end{array}\right)\\
W_{j,\mu} &=& \left(\begin{array}{cc}
\frac{1}{2}W_{j,\mu}^0&\frac{1}{\sqrt{2}}W_{j,\mu}^+\\\frac{1}{\sqrt{2}}W_{j,\mu}^-&-\frac{1}{2}W_{j,\mu}^0
\end{array}\right)\\
W_{n+1,\mu} &=& \left(\begin{array}{cc}
\frac{1}{2}W_{n+1,\mu}^0&0\\0&-\frac{1}{2}W_{n+1,\mu}^0
\end{array}\right)
\end{eqnarray}

The mass matrices of the gauge bosons can be obtained by going to unitary gauge ($\Sigma_j\rightarrow1$).  For the neutral gauge bosons, we find 
\begin{equation}
M_n^2 = \frac{\tilde{g}^2 f^2}{4}\left(\begin{array}{ccccccc}
x^2&-x&0&0&\cdot&0&0\\
-x&2&-1&0&\cdot&0&0\\
0&-1&2&-1&\cdot&0&0\\
\cdot&\cdot&\cdot&\cdot&\cdot&-1&0\\
0&0&0&\cdot&-1&2&-xt\\
0&0&0&\cdot&0&-xt&x^2t^2
\end{array}\right)
\end{equation}
while the matrix $M_\pm^2$ for the charged gauge bosons is $M_n^2$ with the last row and column removed.

The photon is massless and given by the wavefunction
\begin{equation}
v_\gamma = \frac{e}{\tilde{g}}\left(\frac{1}{x},1,\cdots,1,\frac{1}{xt}\right)
\end{equation}
where
\begin{equation}
\frac{1}{e^2} = \frac{1}{g^2} + \frac{n}{\tilde{g}^2} + \frac{1}{g'^2}.
\end{equation}

After diagonalizing the gauge boson mass matrices, we find that the other masses and wavefunctions are given, at leading order in $x$, by the following expressions.  The mass and wavefunction of the light $W$ boson are
\begin{eqnarray}
M_{W0}&=&\frac{\tilde{g}\, f x}{2 \sqrt{(n+1)}}\\
v_{W0}^0&=&1\\
v_{W0}^j&=&\frac{n-j+1}{n+1}x
\end{eqnarray}
where the superscript $0$ refers to the left-most $SU(2)$ group on the moose while the superscript $j = [1...n]$ refers to the $SU(2)$ gauge groups on the interior of the moose.  The masses and wavefunctions of the charged KK modes are
\begin{eqnarray}
M_{Wk}&=&\frac{\tilde{g} f}{\sqrt{2}}\sqrt{1-\cos\left[\frac{k\pi}{n+1}\right]}\label{M_Wk}\\
v_{Wk}^0&=&\frac{-x}{\sqrt{2(n+1)}}\cot\left[\frac{k\pi}{2(n+1)}\right] \\
v_{W_k}^j &=& \sqrt{\frac{2}{n+1}} \sin\left[\frac{jk\pi}{n+1}\right].
\end{eqnarray}
Likewise, the mass and wavefunction of the light $Z$ boson are
\begin{eqnarray}
M_{Z0}&=&\frac{\tilde{g}\,fx}{2c\sqrt{(n+1)}}\\
v_{Z0}^0&=&c\\
v_{Z0}^j&=&\frac{c(n+1)-j/c}{n+1}x\\
v_{Z0}^{n+1}&=&-s,
\end{eqnarray}
where superscript $n+1$ refers to the $U(1)$ group.   The masses and wavefunctions of the neutral KK modes are
\begin{eqnarray}
M_{Zk}&=&\frac{\tilde{g} f}{\sqrt{2}}\sqrt{1-\cos\left[\frac{k\pi}{n+1}\right]}=M_{Wk}\\
v_{Zk}^0&=&\frac{-x}{\sqrt{2(n+1)}}\cot\left[\frac{k\pi}{2(n+1)}\right]\\
v_{Zk}^j&=&\sqrt{\frac{2}{n+1}}\sin\left[\frac{jk\pi}{n+1}\right]\\
v_{Zk}^{n+1}&=&\sqrt{\frac{2}{n+1}}\frac{(-1)^k\, x}{t}\left[(n+1)a_1+b_1\right]\\
&a_1&=\frac{(-1)^k}{4(n+1)}\csc^2\left[\frac{k\pi}{2(n+1)}\right]\\
&&\times\left[(-1)^k\sin\left(\frac{k\pi}{n+1}\right)-t^2\sin\left(\frac{kn\pi}{n+1}\right)\right]\nonumber\\
&b_1&=\frac{-1}{2}\cot\left[\frac{k\pi}{2(n+1)}\right].
\end{eqnarray}

We note that the $W$ gauge boson mass is given by
\begin{equation}
M_W = M_{W0} \equiv \frac{gf}{2\sqrt{n+1}} = \frac{gv}{2}~,
\end{equation}
and, hence, we have the relation
\begin{equation}\label{f v}
f = \sqrt{n+1}\ v~.
\end{equation}
The ratio of the $W$ and $Z$ mass is
\begin{equation}
\frac{M_W}{M_Z} = \frac{M_{W0}}{M_{Z0}} = \frac{1}{c} 
\end{equation}
identifying $c$ with $\cos\theta_W$ at leading order in $x$.

The ratio of $M_W$ to the mass of the first KK mode $M_{W1}$ is
\begin{equation}\label{x def M_W/M_W1}
\frac{M_W}{M_{W_1}} = \frac{x}{\sqrt{2(n+1)\left(1-\cos\left[\frac{\pi}{n+1}\right]\right)}}
\end{equation}
which relates $x$ to the mass ratio $M_W/M_{W_1}$ for a given $n$ at leading order.  From this 
we see  that expansion in $x$ is justified as long as $M_{W_1}\gg M_W$.

\subsection{\label{Fermion Sector}Fermion Sector}
The vertical lines in Figure \ref{n-moose} represent the fermionic fields in the theory.  The vertical lines below the circles represent the left chiral fermions while the vertical lines above the circles are the right chiral fermions.  Each fermion is in a fundamental representation of the gauge group to which it is attached and a singlet under all the other gauge groups except $U(1)_{n+1}$.  The charges under $U(1)_{n+1}$ are as follows:  If the fermion is attached to an $SU(2)$ then its charge is $1/3$ for quarks and $-1$ for leptons.  If the fermion is attached to $U(1)_{n+1}$ its charge is twice its  electromagnetic charge: $0$ for neutrinos, $-2$ for charged leptons, $4/3$ for up type quarks and $-2/3$ for down type quarks.

The fermions attached to the internal sites ($1\leq j\leq n$) are vectorially coupled and are, thus, allowed Dirac masses.  We take these masses to be common, and denote them by $M_F$.  The symmetries also allow Yukawa couplings of fermions at adjacent sites using the nonlinear sigma fields.  We have assumed a very simple form for these couplings, inspired by an extra dimension \cite{Hill:2002me} and represented by the diagonal lines in Figure \ref{n-moose}.   For simplicity, we take
the mass parameter for all the diagonal Yukawa links  -- {\it except} for the two at the
ends of the diagram -- to be $M_F$, the same as the Dirac mass, 
corresponding to a massless fermion  in a five-dimensional model.  The Yukawa links on the ends are taken to be suppressed by factors of $\epsilon_L$ on the left end and $\epsilon_R$ on the right end.  All together, the masses of the fermions and the leading order interactions of the fermions and nonlinear sigma fields are given by
\begin{eqnarray}
\mathcal{L}_{\psi\Sigma} &=& - M_F\Big[
\epsilon_L\bar{\psi}_{L0}\Sigma_0\psi_{R1} - \sum_j\bar{\psi}_{Lj}\psi_{Rj} \label{eq:Yukawa}\\
&+& \sum_j\bar{\psi}_{Lj}\Sigma_j\psi_{R,j+1}+ \bar{\psi}_{Ln}\epsilon_R\Sigma_n\psi_{R,n+1}+h.c.\Big]\nonumber
\label{eq:fifthre}
\end{eqnarray}
where the value of $\epsilon_L$ is the same for all fermions, while $\epsilon_R$ is a diagonal matrix which distinguishes flavors \cite{SekharChivukula:2006cg,Coleppa:2006fu}.  For example for the top and bottom quark we have
\begin{equation}
\epsilon_R = \left(\begin{array}{cc}\epsilon_{Rt}&0\\0&\epsilon_{Rb}\end{array}\right)
\end{equation}

The fermion mass matrix can be diagonalized by performing
unitary transformations on the left- and right-handed fermions separately.  To leading
order in $\epsilon_{L,R}$ we find the following masses and wavefunctions for the lightest fermion, $F_0$, in a given tower (which we associate with an ordinary standard model fermion)
\begin{eqnarray}
M_{F_0}&=&M_F\,\epsilon_L\epsilon_{R_f}\\
v_{LF_0}^0&=&1 \label{eq:lightprof1}\\
v_{LF_0}^j&=&\epsilon_L\\
v_{RF_0}^j&=&\epsilon_{R_f}\\
v_{RF_0}^{n+1}&=&1
\label{eq:lightprof4}
\end{eqnarray}
while the expressions for the heavier states, $F_k$, are
\begin{eqnarray}
M_{F_k}&=&2M_F\cos\left[\frac{(n-k+1)\pi}{2n+1}\right]\label{M_fk}\\
v_{LF_k}^0&=&\frac{\epsilon_L}{\sqrt{2n+1}}\tan\left[\frac{(n-k+1)\pi}{2n+1}\right]\\
v_{LF_k}^j&=&\frac{2(-1)^j}{\sqrt{2n+1}}\sin\left[\frac{2j(n-k+1)\pi}{2n+1}\right]\\
v_{RF_k}^j&=&\frac{(-1)^{n+k+j+1}2}{\sqrt{2n+1}}\sin\left[\frac{2(n-j+1)(n-k+1)\pi}{2n+1}\right]\nonumber\\\\
v_{RF_k}^{n+1}&=&\frac{(-1)^{k}\epsilon_{R_f}}{\sqrt{2n+1}}\tan\left[\frac{(n-k+1)\pi}{2n+1}\right]
\end{eqnarray}
For small $\epsilon_L$, we see that the left-handed component of the lightest fermion in each tower
is primarily located at site 0 -- and the flavor-universal factor $\epsilon_L$ controls the amount
of fermion ``delocalization" along the moose.  Likewise, the right-handed component is primarily located at site $n+1$,  and the flavor-dependent quantities  $\epsilon_{R_f}$ control the degree of delocalization.
Since the amplitude for $t\bar{t} \to W^+ W^-$ scattering will depend on the values of $\epsilon_L$ and $\epsilon_{R_t}$, we need to evaluate these quantities; we will start with $\epsilon_L$ and then use it to constrain $\epsilon_{R_t}$.

Precision electroweak corrections provide a useful source of constraints on the parameters of Higgsless models. While custodial symmetry generally keeps  the tree-level value of $\Delta\rho = \alpha T$ sufficiently small, satisfying the bounds on $S$ at tree level requires some degree of fermion delocalization \cite{Anichini:1994xx,Cacciapaglia:2004rb,Cacciapaglia:2005pa,Foadi:2004ps,Foadi:2005hz,Chivukula:2005bn,Casalbuoni:2005rs,SekharChivukula:2005xm}. In a general Higgsless model, one can calculate the ``ideal delocalization" profile of a fermion along the moose that guarantees $S$ and other precision corrections will vanish at tree level.  However, the $n(+2)$-site model studied here and in  \cite{Coleppa:2006fu}  has been simplified such that the light fermion profile is strictly flat on the interior of the moose ({\it c.f.} Eqns. (\ref{eq:lightprof1}) - (\ref{eq:lightprof4})), rather than being ``ideal".  We therefore quantify the relationship between delocalization ($\epsilon_L$) and $S$ in this model by studying a particular experimental observable.

The coupling $g_{We\nu}$ between the $W$, electron and electron-neutrino is well-measured and lies  close to the SM value.  One may parameterize the deviation in this coupling from the SM value as 
\begin{equation}
g_{We\nu} = g_{We\nu_{SM}}\left(1 + aS + bT + cU\right)
\end{equation}
where $a$, $b$ and $c$ are $O(\alpha)$ parameters.  We have already noted that custodial symmetry makes $T$ small in this model and $U$ is generally suppressed relative to both $S$ and $T$.  Hence, the largest corrections are due to $S$:
\begin{equation}
g_{We\nu} \simeq g_{We\nu_{SM}}\left(1 + aS\right).
\end{equation}
We can ensure $S\simeq0$ at tree level by requiring $g_{We\nu}$ in the $n(+2)$-site model to be the same as in the standard model.    An explicit calculation of $g_{We\nu}$ in this model, which requires expanding the wavefunctions, masses, and couplings to order $\epsilon_L^2$ and order $x^2$,  yields \cite{Coleppa:2006fu} 
\begin{equation}
g_{We\nu_n} = g_{We\nu_{SM}} \left( 1 + \frac{n(n+2)}{6(n+1)}x^2-\frac{n}{2}\epsilon_L^2 \right) .
\end{equation}
Therefore, the condition
\begin{equation}
\epsilon_L^2 = \frac{n+2}{3(n+1)}x^2
\end{equation}
causes $S$ to vanish at tree-level. Using  Eqn. (\ref{x def M_W/M_W1}) this
is equivalent to 
\begin{equation}
\epsilon_L^2 =  \frac{2}{3}(n+2)\left(1-\cos\left[\frac{\pi}{n+1}\right]\right)\frac{M_W^2}{M_{W_1}^2}~,
\end{equation}
in terms of physical masses.
Here again, note that  $\epsilon_L$ is small so long as $M_W\ll M_{W_1}$.

Finally, the parameter $\epsilon_{R_f}$ can be determined by taking the ratio of the masses of the light fermion and the first KK mode.
\begin{equation}
\frac{M_{F_0}}{M_{F_1}} = \frac{\epsilon_L\epsilon_{R_f}}{2\cos\left[\frac{n\pi}{2n+1}\right]}
\end{equation}
Since we know $\epsilon_L$, this gives a prediction for $\epsilon_{R_f}$ in terms of physical masses
\begin{equation}
\epsilon_{R_f} = \frac{\sqrt{6}\cos\left[\frac{n\pi}{2n+1}\right]}{\sqrt{(n+2)\left(1-\cos\left[\frac{\pi}{n+1}\right]\right)}}\frac{M_{F_0}}{M_{F_1}}\frac{M_{W_1}}{M_W}~.
\end{equation}
For all flavors except the top quark, this parameter is tiny; at leading order, we therefore set $\epsilon_{R_f}=0$ for all the light fermions.  The size of $\epsilon_{R_t}$ affects $\Delta\rho$ at one loop; comparison of the experimental bounds on $\Delta\rho$ with the value calculated in Higgsless models  \cite{SekharChivukula:2006cg,Coleppa:2006fu}  shows that $\epsilon_{R_t}$ must also be relatively small.  In what follows, we therefore keep only the leading terms in $\epsilon_{R_t}$.

\subsection{\label{Goldstone Boson Sector}Goldstone Boson Sector}

We will perform the computation of  the process $t_+ \bar{t}_+ \to W^+_L W^-_L$ in the $n(+2)$
site model using the equivalence theorem. We must, therefore, 
determine the wavefunction of the Goldstone bosons associated with (eaten by) the massive gauge bosons.  This is determined by the mixing between the two given in Eqn. (\ref{L_D Sigma}).  To find the mixing, we expand the nonlinear sigma-model field $\Sigma_j$ and keep the terms linear in both the gauge bosons ($W_j$) and the Goldstone bosons ($\pi_j$).  After these manipulations, Eqn. (\ref{L_D Sigma}) becomes 
\begin{eqnarray}
\mathcal{L}_{\pi W} = -i\frac{\tilde{g}f}{2} \Bigg[
&\Big\{\partial_\mu\pi_0\ ,\ x\ W_0^\mu-W_1^\mu\Big\}&\\
&+\sum_{j=1}^{n-1} \Big\{\partial_\mu\pi_j\ ,\ W_j^\mu-W_{j+1}^\mu\Big\}&\nonumber\\
&+\Big\{\partial_\mu\pi_n\ ,\ W_n^\mu-xt\ W_{n+1}^\mu\Big\}&
\Bigg]\nonumber
\end{eqnarray}
from which we may read off the wavefunctions for the charged Goldstone bosons as
\begin{eqnarray}
v_{\pi_k^\pm}^{[0]} &=& \frac{1}{N_{\pi_k^\pm}}\left(x\ v_{W_k}^0 - v_{W_k}^1 \right)\\
v_{\pi_k^\pm}^{[j]} &=& \frac{1}{N_{\pi_k^\pm}}\left( v_{W_k}^j - v_{W_k}^{j+1} \right)\\
v_{\pi_k^\pm}^{[n]} &=& \frac{1}{N_{\pi_k^\pm}}\  v_{W_k}^n
\end{eqnarray}
where the $N_{\pi_k}$ are normalization factors.  Note that Nambu-Goldstone boson components are associated with the links rather than the gauge groups: the superscript [0] refers to the left-most link, the superscript [n] refers to the right-most link, and the superscripts [j] range from 1 through n-1 and denote the interior links of the Moose. The wavefunctions for the neutral Goldstone bosons are similar
\begin{eqnarray}
v_{\pi_k^0}^{[0]} &=& \frac{1}{N_{\pi_k^0}}\left(x\ v_{Z_k}^0 - v_{Z_k}^1 \right)\\
v_{\pi_k^0}^{[j]} &=& \frac{1}{N_{\pi_k^0}}\left( v_{Z_k}^j - v_{Z_k}^{j+1} \right)\\
v_{\pi_k^0}^{[n]} &=& \frac{1}{N_{\pi_k^0}}\left( v_{Z_k}^n - xt\ v_{Z_k}^{n+1} \right)~,
\end{eqnarray}
but include a contribution from the $Z_k$ wavefunction on the $U(1)$ site.  

These wavefunctions are particularly simple for the lightest
modes, the  $W$ and $Z$: they are flat
\begin{equation}
v_{\pi_0^\pm}^{[l]} = \frac{1}{\sqrt{n+1}} = 
v_{\pi_0^0}^{[l]} 
\end{equation}
with the same value on all links [$l = 0...n$] of the Moose.

\subsection{\label{Couplings}Couplings}

To obtain the couplings of the Goldstone bosons to the fermions, we start from Eqn. (\ref{eq:fifthre}),  expand the nonlinear sigma-model  fields,  and plug in the eigenmode wavefunctions we have just derived.  Doing this, we find
\begin{eqnarray}
g_{LtF_k\pi} &=& -i\frac{\sqrt{2} M_F}{f}\bigg[\epsilon_Lv_{Lt}^0v_{RF_k}^1v_\pi^{[0]} +\sum_i v_{Lt}^iv_{RF_k}^{i+1}v_\pi^{[i]}\nonumber\\
&&\hspace{0.6in} +\epsilon_{Rb}v_{Lt}^nv_{RF_k}^{n+1}v_\pi^{[n]}\bigg]
\nonumber\\
  && \hspace{-0.5in} =  (-1)^k \frac{i\sqrt{2}M_F\epsilon_L}{\sqrt{2n+1}(n+1)v}\tan\left[\frac{(n-k+1)\pi}{2n+1}\right]
\label{eq:gLcoupling}
\end{eqnarray}
\begin{eqnarray}
g_{RtF_k\pi} &=& -i\frac{\sqrt{2}M_F}{f}\bigg[\epsilon_Lv_{LF_k}^0v_{Rt}^1v_\pi^{[0]} +\sum_i v_{LF_k}^i v_{Rt}^{i+1} v_\pi^{[i]}\nonumber\\
&&\hspace{0.6in} +\epsilon_{Rt}v_{LF_k}^nv_{Rt}^{n+1}v_\pi^{[n]}\bigg]\nonumber\\
&=& \frac{i\sqrt{2}M_F\epsilon_R}{\sqrt{2n+1}(n+1)v}\tan\left[\frac{(n-k+1)\pi}{2n+1}\right]
\end{eqnarray}
\begin{eqnarray}
g_{tt\pi^+\pi^-} &=& \frac{M_F}{f^2}\bigg[\epsilon_Lv_{Lt}^0v_{Rt}^1(v_\pi^{[0]})^2 + \sum_i v_{Lt}^i v_{Rt}^{i+1}(v_\pi^{[i]})^2 \nonumber\\
&&\hspace{0.6in} + \epsilon_{Rt}v_{Lt}^nv_{Rt}^{n+1}(v_\pi^{[n]})^2\bigg]
\nonumber\\
&=& \frac{m_t}{(n+1)v^2}~.\label{eq:g4coupling}
\end{eqnarray}
Here we have denoted the lightest fermions (previously denoted $F_0$) by $t$ and $b$, as appropriate, 
while leaving the corresponding
$KK$ modes as $F_k$ (which, to leading order in $\epsilon_{L,R}$, have the same properties
for all quarks).
Note that  the four point vertex has an extremely simple form, and vanishes in
the limit $n \to \infty$.

\section{Unitarity Bounds on $t\bar{t}\to W_L W_L$}

\begin{figure}
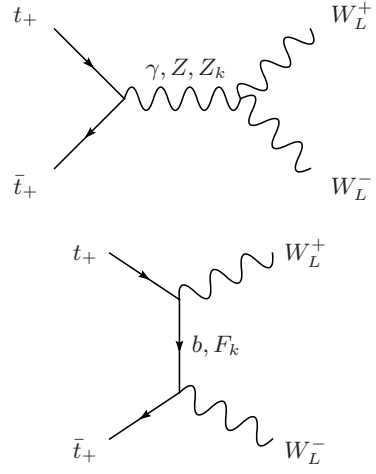

  \begin{center}
    \includegraphics{images/2f-2G.eps}
    \begin{picture}(0,0)(0,0)
      \put(-120,60){$t_+$}
      \put(-120,-5){$\bar{t}_+$}
      \put(0,60){$W_L^+$}
      \put(0,-5){$W_L^-$}
      \put(-70,40){$\gamma,Z,Z_k$}
    \end{picture}
    \vspace{0.3in}\\
    \includegraphics[]{images/2f-nu-2G.eps}
    \begin{picture}(0,0)(0,0)
      \put(-80,75){$t_+$}
      \put(-80,0){$\bar{t}_+$}
      \put(0,75){$W_L^+$}
      \put(0,0){$W_L^-$}
      \put(-35,40){$b,F_k$}
    \end{picture}
    \caption{\label{ttbar - k - WW}The diagrams that contribute to the process $t_+\bar{t}_+\rightarrow W_L^+W_L^-$ in the $n(+2)$ site Higgsless model.  There are analogous diagrams for the process $t_-\bar{t}_-\rightarrow W_L^+W_L^-$.  As in the SM, most of the linear growth in $\sqrt{s}$ will cancel.  All the persisting linear growth in $\sqrt{s}$ comes from the $t$ channel diagrams.}
  \end{center}
\end{figure}

\begin{figure}
    \includegraphics[]{images/2f--2phi.eps}
    \begin{picture}(0,0)(0,0)
      \put(-85,70){$t_+$}
      \put(-85,-5){$\bar{t}_+$}
      \put(5,70){$\pi^+$}
      \put(5,-5){$\pi^-$}
    \end{picture}
\vspace{0.5in}

\includegraphics{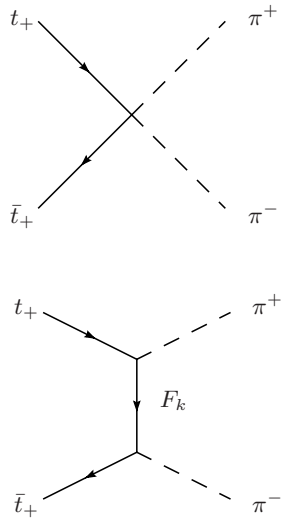}
\begin{picture}(0,0)(0,0)
  \put(-85,70){$t_+$}
  \put(-85,-5){$\bar{t}_+$}
  \put(-30,35){$F_k$}
  \put(5,70){$\pi^+$}
  \put(5,-5){$\pi^-$}
\end{picture}
\caption{\label{n site diagram hel nonconservin}Diagrams contributing to unitarity violation at high energies in the process $t_+\bar{t}_+\rightarrow\pi^+\pi^-$.  There are analogous diagrams for the process $t_-\bar{t}_-\rightarrow\pi^+\pi^-$.  The top diagram grows linearly with $\sqrt{s}$ for all energies, whereas the bottom diagrams only grow with $\sqrt{s}$ up to $M_{F_k}$, after which they fall off as $1/\sqrt{s}$.}
\end{figure}

The diagrams that contribute at tree level to $t_+\bar{t}_+\to W^+_L W^-_L$ are shown in Figure \ref{ttbar - k - WW}.
We are again interested in the behavior at large energies, so we expand in the small parameters $M_W^2/s$ and $m_t^2/s$; we also include all colors and both helicity polarizations in a coupled channel analysis (Eqn. (\ref{in state})).
The calculation is most easily performed using the equivalence theorem \cite{Cornwall:1974km,Vayonakis:1976vz}.
Again, as in the SM (see Figure \protect\ref{AC: 2f-G-2pi}),
the potential $s$-channel diagrams do not contribute to the $J=0$ amplitude, and
the only diagrams that contribute are shown in 
Figure \ref{n site diagram hel nonconservin}.
The scattering amplitude arising from the diagrams in Figure 6 is
\begin{equation}\label{M_11 n}
\mathcal{M}  = \sqrt{6s}\left(g_{tt\pi^+\pi^-} - \sum_k \frac{M_{F_k}g_{LtF_k\pi}\,g_{RtF_k\pi}}{t-M_{F_k}^2}\right)
\end{equation}
where the couplings are given in Eqns. (\ref{eq:gLcoupling}) -- (\ref{eq:g4coupling}).

\begin{figure}
\begin{center}
\includegraphics[scale=0.8]{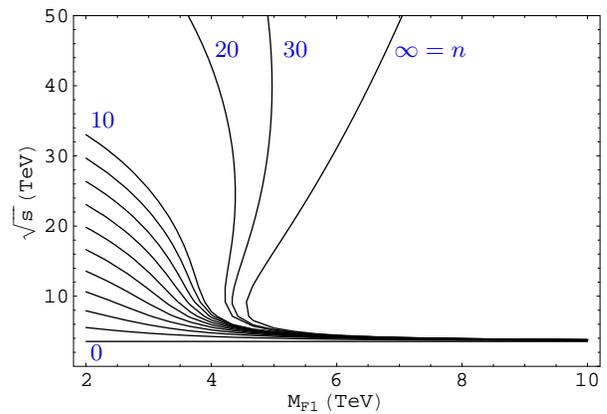}
\begin{picture}(0,0)
\put(-200,20){\textcolor{blue}{$0$}}
\put(-200,108){\textcolor{blue}{$10$}}
\put(-153,135){\textcolor{blue}{$20$}}
\put(-127,135){\textcolor{blue}{$30$}}
\put(-85,135){\textcolor{blue}{$\infty = n$}}
\end{picture}
\end{center}
\caption{\label{nsite_uni_bound_hel_non}The scale where unitarity breaks down in the helicity nonconserving channel in the $n(+2)$ site model.  Unitarity is valid in the region below and to the left
of a given curve.  The bottom-most curve is for $n=0$ and is the AC bound.  The line directly above the bottom one is for $n=1$ and corresponds to the Three Site Model.  The line directly above that is for $n=2$ and so on until $n=10$.  The line above that is for $n=20$, the line to the right of that is for $n=30$ and the line to the right of that is the continuum limit ($n\rightarrow\infty$).  We find that unitarity breaks down if either $E$ is large or $M_{F_1}$ is large.  If $M_{F_1}$ is large, then unitarity breaks down for $\sqrt{s}$ very close to the AC bound.  On the other hand, if $M_{F_1}\lesssim 4.5\,{\rm TeV}$, unitarity can be valid in this process to very high energies, with the precise value depending on the number
of sites $n$.  }
\end{figure}

\begin{figure}
\begin{center}
\includegraphics[scale=0.8]{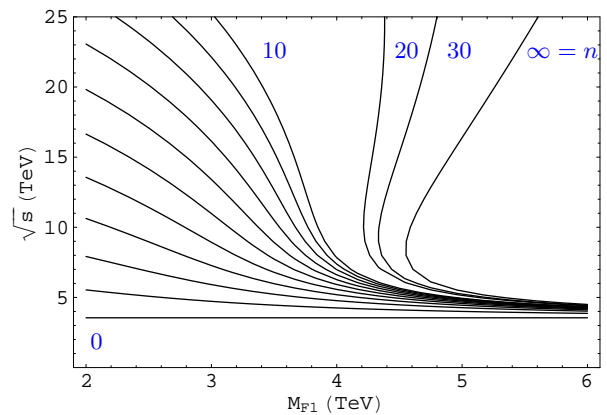}
\begin{picture}(0,0)
\put(-200,25){\textcolor{blue}{$0$}}
\put(-135,135){\textcolor{blue}{$10$}}
\put(-85,135){\textcolor{blue}{$20$}}
\put(-65,135){\textcolor{blue}{$30$}}
\put(-35,135){\textcolor{blue}{$\infty = n$}}
\end{picture}
\end{center}
\caption{\label{nsite_uni_bound_hel_non_blow_up}Expanded view of low $\sqrt{s}$ region  of
Figure \protect\ref{nsite_uni_bound_hel_non}.}
\end{figure}

The $J=0$ partial wave can be extracted as
\begin{eqnarray}
a_0 &=& \frac{1}{32\pi}\int_{-1}^1d\cos\theta \mathcal{M}\\
&=& \frac{\sqrt{6}}{16\pi}\left[
g_{tt\pi^+\pi^-}\sqrt{s} + \sum_k g_{LtF_k\pi}\,g_{RtF_k\pi} \, g\left(\frac{\sqrt{s}}{M_{F_k}}\right)
\right]\nonumber
\end{eqnarray}
where
\begin{equation}
g(x) = \frac{1}{x}\mbox{ln}(1+x^2)
\end{equation}
This partial wave must be less than $1/2$ to maintain unitarity, giving a bound on $\sqrt{s}$ and/or $M_{F_1}$.  We have plotted this bound in Figures  \ref{nsite_uni_bound_hel_non} 
and \ref{nsite_uni_bound_hel_non_blow_up} for $n=0, 1, 2, \cdots , 10, 20, 30$ and $\infty$.
The $n=0$ bound corresponds to the original AC bound of Eqn. (\ref{eq:ouracresult}).

We see from these figures that there are two important domains corresponding to different ranges of values for $M_{F_1}$.  In the first domain, where $M_{F_1}\lesssim 4.5 \,{\rm TeV}$, we find that unitarity can be satisfied up to very large energies.  In this limit, we find that the $t$ channel diagram becomes irrelevant and the process is controlled by the four point vertex (Figure \ref{n site diagram hel nonconservin}).  For the lowest fermion masses, $M_{F_1}\ll 4.5\,{\rm TeV}$, we find
\begin{equation}
a_0 \simeq \frac{\sqrt{6\,s}\,m_t}{16\pi v^2(n+1)} \lesssim \frac{1}{2}
\end{equation}
which gives the bound
\begin{equation}
\sqrt{s} \lesssim (n+1)\,3.5\, {\rm TeV}
\end{equation}
In this ``low" KK  fermion-mass region, unitarity is valid to approximately $(n+1)$ times the AC bound.

In the second domain, where $M_{F_1}>4.5\,{\rm TeV}$, we find that, for all $n$, unitarity breaks down at a value of $\sqrt{s}$ given approximately by the AC bound (Eqn. (\ref{eq:ouracresult}))  In Figures \ref{nsite_uni_bound_hel_non} and \protect\ref{nsite_uni_bound_hel_non_blow_up}, we see that at $M_{F_1} \sim 4.5$ TeV, the curves corresponding to small $n$ approach the $n=0$ curve, while the curves for large $n$ turn back on themselves, defining a 
 wedge-shaped area in which unitarity is always violated starting at $\sqrt{s}$ of order a few TeV.

To understand why $M_{F_1}$ = 4.5 TeV is the fermion mass value at which the theory crosses from the first to the second domain, we consider what happens as $n\rightarrow\infty$.  In this limit, the four point vertex disappears and we are left with the partial wave amplitude
\begin{equation}
\lim_{n\rightarrow\infty}a_0 = 
\frac{2\sqrt{6}M_{F_1}m_t}{\pi^4v^2}
\sum_k \frac{(-1)^{k+1}}{(2k-1)^2}\,
g\left(\frac{\sqrt{s}}{(2k-1)M_{F_1}}\right)~.
\end{equation}
This sum is dominated by the first KK mode ($k=1$).  Thus, to locate the left most edge of the 
wedge-shaped in the $(\sqrt{s}, M_{F1})$ plane where unitarity is violated, we need only keep the first KK fermion mode
\begin{equation}
\lim_{n\rightarrow\infty}a_0(k=1) \approx 
\frac{2\sqrt{6}M_{F_1}m_t}{\pi^4v^2} 
g\left(\frac{\sqrt{s}}{M_{F_1}}\right)~.
\end{equation}
The function $g(\sqrt{s}/M_{F_1})$ determines the shape of this bound. It is maximized for $\sqrt{s}
= 2 M_{F1}$  and gives  the upper limit of $M_{F_1}$,  
\begin{equation}
M_{F_1} \lesssim \frac{\pi^4v^2}{2\sqrt{6}m_t\mbox{ln}(5)} \sim 4.25\, {\rm TeV}~,
\end{equation}
if we want this amplitude to be unitary  up to very high scales.
Including the higher fermion KK modes changes this upper bound only
slightly, to $\sim4.5\,{\rm TeV}$.
Note that,
in the continuum limit, the scattering amplitude does not grow at asymptotically high energies --
a property ensured by various sum-rules satisfied by the couplings
\protect\cite{Schwinn:2004xa,Schwinn:2005qa}. Nonetheless, as illustrated in Figures \protect\ref{nsite_uni_bound_hel_non}
and \protect\ref{nsite_uni_bound_hel_non_blow_up}, the properly normalized spin-0 coupled-channel amplitude 
exceeds the unitarity bound for various ranges of $\sqrt{s}$ and $M_{F1}$.

While our work demonstrates that the bound on the scale
of fermion mass generation is independent of the bound on the scale
of gauge-boson mass generation in these models, the physical significance of
the fermion-mass-generation bound depends on the ``high-energy" (UV) completion which underlies the
$n(+2)$ site model. The simplest possible UV completion is one in which each
of the nonlinear sigma-model link theories is replaced by a linear 
Gell-mann--Levy sigma model. In this case, the strength of the adjacent site  couplings
in Eqn.  (\ref{eq:Yukawa}) is determined by a dimensionless Yukawa coupling
of order $M_F/f$. The large-$M_F$ limit, therefore, corresponds to large Yukawa
coupling. In this case, the bound on $M_F$ is expected to be related to the
triviality bound on the corresponding Yukawa coupling \cite{Chanowitz:1978uj,Chanowitz:1978mv,Einhorn:1986za}.

\section{Summary}

In this paper we have examined upper bounds on the scale of
top-quark mass generation in viable deconstructed Higgsless
models.  These bounds are derived from the scale at which 
unitarity is violated in the helicity nonconserving 
amplitude for top-anti-top pairs to scatter into pairs 
of longitudinally polarized electroweak gauge bosons. 
We have shown that the scale of unitarity violation in this
process depends on the mass of the additional vector-like fermion
states that occur in these theories and, in this sense,
the scale of fermion mass generation is {\it separate} from that
of gauge-boson mass generation.  For sufficiently light vector
fermions, and for a deconstructed theory with sufficiently many lattice
sites (that is, sufficiently close to the continuum limit), we have shown
that the Appelquist-Chanowitz  bound on top-quark mass generation
is substantially weakened,
while the bound is recovered as one increases the mass
of the vector-like fermions. Our results are expected to apply to
any model in which top-quark mass generation occurs, in part, through
mixing between chiral and vector fermions.

\section{Acknowledgements}

This work was supported in part by the US National Science Foundation under
grant  PHY-0354226. We thank Stefano DiChiara and Hong-Jian He for useful conversations.


\begin{thebibliography}{99}

\bibitem{Higgs:1964ia}
  P.~W.~Higgs,
  Phys.\ Lett.\  {\bf 12}, 132 (1964).

\bibitem{Weinberg:1967tq}
  S.~Weinberg,
  Phys.\ Rev.\ Lett.\  {\bf 19}, 1264 (1967).

\bibitem{Salam:1968rm}
  A.~Salam, 1968.

\bibitem{Appelquist:1987cf}
  T.~Appelquist and M.~S.~Chanowitz,
  {\it Unitarity  Bound on the Scale of Fermion Mass Generation},
  Phys.\ Rev.\ Lett.\  {\bf 59}, 2405 (1987)
  [Erratum-ibid.\  {\bf 60}, 1589 (1988)].
  
\bibitem{Marciano:1989ns}
  W.~J.~Marciano, G.~Valencia and S.~Willenbrock,
  Phys.\ Rev.\  D {\bf 40}, 1725 (1989).
  
\bibitem{Golden:1994pj}
  M.~Golden,
  Phys.\ Lett.\  B {\bf 338}, 295 (1994)
  [arXiv:hep-ph/9408272].


\bibitem{Maltoni:2001dc}
  F.~Maltoni, J.~M.~Niczyporuk and S.~Willenbrock,
  Phys.\ Rev.\  D {\bf 65}, 033004 (2002)
  [arXiv:hep-ph/0106281].

\bibitem{Dicus:2004rg}
  D.~A.~Dicus and H.~J.~He,
  Phys.\ Rev.\  D {\bf 71}, 093009 (2005)
  [arXiv:hep-ph/0409131].
  


\bibitem{LlewellynSmith:1973ey}
  C.~H.~Llewellyn Smith,
  Phys.\ Lett.\  B {\bf 46}, 233 (1973).

\bibitem{Dicus:1992vj}
  D.~A.~Dicus and V.~S.~Mathur,
  Phys.\ Rev.\  D {\bf 7}, 3111 (1973).

\bibitem{Cornwall:1974km}
  J.~M.~Cornwall, D.~N.~Levin and G.~Tiktopoulos,
  Phys.\ Rev.\  D {\bf 10}, 1145 (1974)
  [Erratum-ibid.\  D {\bf 11}, 972 (1975)].

\bibitem{Lee:1977eg}
  B.~W.~Lee, C.~Quigg and H.~B.~Thacker,
  Phys.\ Rev.\  D {\bf 16}, 1519 (1977).
  
\bibitem{Veltman:1976rt}
  M.~J.~G.~Veltman,
  Acta Phys.\ Polon.\  B {\bf 8}, 475 (1977).

\bibitem{Csaki:2003dt}
  C.~Csaki, C.~Grojean, H.~Murayama, L.~Pilo and J.~Terning,
{\it Gauge theories on an interval: Unitarity without a Higgs},
  Phys.\ Rev.\ D {\bf 69}, 055006 (2004)
  [arXiv:hep-ph/0305237].



 \bibitem{SekharChivukula:2001hz}
R.~Sekhar~Chivukula, D.~A. Dicus, and H.-J. He, {\it Unitarity of compactified
  five dimensional yang-mills theory},  {\em Phys. Lett.} {\bf B525} (2002)
  175--182, [arXiv:hep-ph/0111016].

\bibitem{Chivukula:2002ej}
R.~S. Chivukula and H.-J. He, {\it Unitarity of deconstructed five-dimensional
  yang-mills theory},  {\em Phys. Lett.} {\bf B532} (2002) 121--128,
  [arXiv:hep-ph/0201164].

\bibitem{Chivukula:2003kq}
R.~S. Chivukula, D.~A. Dicus, H.-J. He, and S.~Nandi, {\it Unitarity of the
  higher dimensional standard model},  {\em Phys. Lett.} {\bf B562} (2003)
  109--117, [arXiv:hep-ph/0302263].

\bibitem{He:2004zr}
H.-J.~He,
{\it Higgsless deconstruction without boundary condition},
arXiv:hep-ph/0412113.

\bibitem{Antoniadis:1990ew}
  I.~Antoniadis,
  Phys.\ Lett.\ B {\bf 246}, 377 (1990).



%
\bibitem{Agashe:2003zs}
  K.~Agashe, A.~Delgado, M.~J.~May and R.~Sundrum,
  {\it RS1, Custodial Isospin and Precision Tests},
  JHEP {\bf 0308}, 050 (2003)
  [arXiv:hep-ph/0308036].

\bibitem{Csaki:2003zu}
C.~Csaki, C.~Grojean, L.~Pilo, and J.~Terning, {\it Towards a realistic model
  of higgsless electroweak symmetry breaking},  {\em Phys. Rev. Lett.} {\bf 92}
  (2004) 101802, [arXiv:hep-ph/0308038].

\bibitem{Burdman:2003ya}
  G.~Burdman and Y.~Nomura,
 {\it Holographic theories of electroweak symmetry breaking without a Higgs
  boson},
  Phys.\ Rev.\ D {\bf 69}, 115013 (2004)
  [arXiv:hep-ph/0312247].

\bibitem{Cacciapaglia:2004jz}
  G.~Cacciapaglia, C.~Csaki, C.~Grojean and J.~Terning,
  {\it Oblique corrections from Higgsless models in warped space},
  {\em Phys. Rev. D} {\bf 70}, (2004) 075014,
  [arXiv:hep-ph/0401160].

\bibitem{Arkani-Hamed:2001ca}
N.~Arkani-Hamed, A.~G. Cohen, and H.~Georgi, {\it (de)constructing dimensions},
   {\em Phys. Rev. Lett.} {\bf 86} (2001) 4757--4761,
  [arXiv:hep-th/0104005].

\bibitem{Hill:2000mu}
C.~T. Hill, S.~Pokorski, and J.~Wang, {\it Gauge invariant effective lagrangian
  for kaluza-klein modes},  {\em Phys. Rev.} {\bf D64} (2001) 105005,
  [arXiv:hep-th/0104035].
\bibitem{Foadi:2003xa}
R.~Foadi, S.~Gopalakrishna, and C.~Schmidt, {\it Higgsless electroweak symmetry
  breaking from theory space},  {\em JHEP} {\bf 03} (2004) 042,
  [arXiv: hep-ph/0312324].

\bibitem{Hirn:2004ze}
  J.~Hirn and J.~Stern,
  {\it The role of spurions in Higgs-less electroweak effective theories},
  Eur.\ Phys.\ J.\ C {\bf 34}, 447 (2004)
  [arXiv:hep-ph/0401032].

\bibitem{Casalbuoni:2004id}
R.~Casalbuoni, S.~De Curtis and D.~Dominici,
{\it Moose models with vanishing S parameter},
Phys.\ Rev.\ D {\bf 70} (2004) 055010
[arXiv:hep-ph/0405188].

\bibitem{Chivukula:2004pk}
R.~S.~Chivukula, E.~H.~Simmons, H.~J.~He, M.~Kurachi and M.~Tanabashi,
{\it The structure of corrections to electroweak interactions in Higgsless
models},
Phys.\ Rev.\ D {\bf 70} (2004) 075008
[arXiv:hep-ph/0406077].


\bibitem{Perelstein:2004sc}
M.~Perelstein, {\it Gauge-assisted technicolor?},  {\em JHEP} {\bf 10} (2004)
  010, [arXiv:hep-ph/0408072].

\bibitem{Georgi:2004iy}
  H.~Georgi, {\it Fun with Higgsless theories},
  Phys.\ Rev.\ D {\bf 71}, 015016 (2005)
  [arXiv:hep-ph/0408067].


\bibitem{SekharChivukula:2004mu}
R.~Sekhar Chivukula, E.~H.~Simmons, H.~J.~He, M.~Kurachi and M.~Tanabashi,
{\it Electroweak corrections and unitarity in linear moose models},
Phys.\ Rev.\ D {\bf 71} (2005) 035007
[arXiv:hep-ph/0410154].


\bibitem{SekharChivukula:2006cg}
  R.~Sekhar Chivukula, B.~Coleppa, S.~Di Chiara, E.~H.~Simmons, H.~J.~He, M.~Kurachi and M.~Tanabashi,
  ``A three site higgsless model,''
  arXiv:hep-ph/0607124.

\bibitem{Casalbuoni:1985kq}
  R.~Casalbuoni, S.~De Curtis, D.~Dominici and R.~Gatto,
  Phys.\ Lett.\  B {\bf 155}, 95 (1985).

\bibitem{Anichini:1994xx}
  L.~Anichini, R.~Casalbuoni and S.~De Curtis,
  Phys.\ Lett.\  B {\bf 348}, 521 (1995)
  [arXiv:hep-ph/9410377].


\bibitem{Cacciapaglia:2004rb}
G.~Cacciapaglia, C.~Csaki, C.~Grojean and J.~Terning,
{\it Curing the ills of Higgsless models: The S parameter and unitarity},
Phys.\ Rev.\ D {\bf 71} (2005) 035015
[arXiv:hep-ph/0409126].

\bibitem{Cacciapaglia:2005pa}
  G.~Cacciapaglia, C.~Csaki, C.~Grojean, M.~Reece and J.~Terning,
 {\it Top and bottom: A brane of their own},
  {\em Phys.  Rev.  D}  {\bf 72}, (2005) 095018
  [arXiv:hep-ph/0505001].

\bibitem{Foadi:2004ps}
R.~Foadi, S.~Gopalakrishna and C.~Schmidt,
{\it Effects of fermion localization in Higgsless theories and electroweak
constraints},
Phys.\ Lett.\ B {\bf 606} (2005) 157
[arXiv:hep-ph/0409266].

\bibitem{Foadi:2005hz}
  R.~Foadi and C.~Schmidt,
 {\it An Effective Higgsless Theory: Satisfying Electroweak Constraints and a
  Heavy Top Quark},
  Phys.\ Rev.\ D {\bf 73} (2006)  075011
  [arXiv:hep-ph/0509071].

\bibitem{Chivukula:2005bn}
  R.~S.~Chivukula, E.~H.~Simmons, H.~J.~He, M.~Kurachi and M.~Tanabashi,
   {\it Deconstructed Higgsless models with one-site delocalization},
  %
  Phys.\ Rev.\ D {\bf 71}, 115001 (2005)
  [arXiv:hep-ph/0502162].

\bibitem{Casalbuoni:2005rs}
R.~Casalbuoni, S.~De Curtis, D.~Dolce and D.~Dominici,
{\it Playing with fermion couplings in Higgsless models},
Phys.\ Rev.\ D {\bf 71}, 075015 (2005)
[arXiv:hep-ph/0502209].


\bibitem{SekharChivukula:2005xm}
  R.~Sekhar Chivukula, E.~H.~Simmons, H.~J.~He, M.~Kurachi and M.~Tanabashi,
  {\it Ideal fermion delocalization in Higgsless models},
    Phys.\ Rev.\ D {\bf 72}, 015008 (2005)
  [arXiv:hep-ph/0504114].


\bibitem{Peskin:1992sw}
M.~E. Peskin and T.~Takeuchi, {\it Estimation of oblique electroweak
  corrections},  {\em Phys. Rev.} {\bf D46} (1992) 381--409.

\bibitem{Altarelli:1990zd}
G.~Altarelli and R.~Barbieri, {\it Vacuum polarization effects of new physics
  on electroweak processes},  {\em Phys. Lett.} {\bf B253} (1991) 161--167.

\bibitem{Altarelli:1991fk}
G.~Altarelli, R.~Barbieri, and S.~Jadach, {\it Toward a model independent
  analysis of electroweak data},  {\em Nucl. Phys.} {\bf B369} (1992) 3--32.

\bibitem{Barbieri:2004qk}
  R.~Barbieri, A.~Pomarol, R.~Rattazzi and A.~Strumia,
 {\it Electroweak symmetry breaking after LEP1 and LEP2},
  Nucl.\ Phys.\ B {\bf 703}, 127 (2004)
  [arXiv:hep-ph/0405040].
  
  \bibitem{Chivukula:2004af}
R.~S. Chivukula, E.~H. Simmons, H.-J. He, M.~Kurachi, and M.~Tanabashi, {\it
  Universal non-oblique corrections in higgsless models and beyond},  {\em
  Phys. Lett.} {\bf B603} (2004) 210--218,
  [arXiv:hep-ph/0408262].

\bibitem{SekharChivukula:2006we}
  R.~Sekhar Chivukula, E.~H.~Simmons, H.~J.~He, M.~Kurachi and M.~Tanabashi,
  Phys.\ Rev.\  D {\bf 75}, 035005 (2007)
  [arXiv:hep-ph/0612070].

\bibitem{Coleppa:2006fu}
  B.~Coleppa, S.~Di Chiara and R.~Foadi,
  {\it One loop corrections to the rho parameter in Higgsless models},
  arXiv:hep-ph/0612213.
  
\bibitem{Dobrescu:1997nm}
  B.~A.~Dobrescu and C.~T.~Hill,
 {\it Electroweak symmetry breaking via top condensation seesaw},
  Phys.\ Rev.\ Lett.\  {\bf 81}, 2634 (1998)
  [arXiv:hep-ph/9712319].
  
\bibitem{Chivukula:1998wd}
  R.~S.~Chivukula, B.~A.~Dobrescu, H.~Georgi and C.~T.~Hill,
 {\it Top quark seesaw theory of electroweak symmetry breaking},
  Phys.\ Rev.\  D {\bf 59}, 075003 (1999)
  [arXiv:hep-ph/9809470].

\bibitem{He:2001fz}
  H.~J.~He, C.~T.~Hill and T.~M.~P.~Tait,
 {\it Top quark seesaw, vacuum structure and electroweak precision
  constraints},
  Phys.\ Rev.\  D {\bf 65}, 055006 (2002)
  [arXiv:hep-ph/0108041].

\bibitem{Suzuki:1991kh}
  M.~Suzuki,
  Phys.\ Rev.\  D {\bf 44}, 3628 (1991).

\bibitem{Lebed:1991qv}
  R.~F.~Lebed and M.~Suzuki,
  Phys.\ Rev.\  D {\bf 45}, 1744 (1992).

\bibitem{Kaplan:1991dc}
  D.~B.~Kaplan,
  Nucl.\ Phys.\  B {\bf 365}, 259 (1991).

\bibitem{Maldacena:1998re}
J.~M. Maldacena, {\it The large n limit of superconformal field theories and
  supergravity},  {\em Adv. Theor. Math. Phys.} {\bf 2} (1998) 231--252,
  [\href{http://xxx.lanl.gov/abs/hep-th/9711200}{{\tt hep-th/9711200}}].

\bibitem{Gubser:1998bc}
S.~S. Gubser, I.~R. Klebanov, and A.~M. Polyakov, {\it Gauge theory correlators
  from non-critical string theory},  {\em Phys. Lett.} {\bf B428} (1998)
  105--114, [\href{http://xxx.lanl.gov/abs/hep-th/9802109}{{\tt
  hep-th/9802109}}].

\bibitem{Witten:1998qj}
E.~Witten, {\it Anti-de sitter space and holography},  {\em Adv. Theor. Math.
  Phys.} {\bf 2} (1998) 253--291,
  [\href{http://xxx.lanl.gov/abs/hep-th/9802150}{{\tt hep-th/9802150}}].

\bibitem{Aharony:1999ti}
O.~Aharony, S.~S. Gubser, J.~M. Maldacena, H.~Ooguri, and Y.~Oz, {\it Large n
  field theories, string theory and gravity},  {\em Phys. Rept.} {\bf 323}
  (2000) 183--386, [\href{http://xxx.lanl.gov/abs/hep-th/9905111}{{\tt
  hep-th/9905111}}].

\bibitem{Csaki:2005vy}
  C.~Csaki, J.~Hubisz and P.~Meade,
  arXiv:hep-ph/0510275.



\bibitem{Appelquist:1980ae}
  T.~Appelquist and C.~W.~Bernard,
   {\it The Nonlinear Sigma Model In The Loop Expansion},
  %
  Phys.\ Rev.\ D {\bf 23}, 425 (1981).


\bibitem{Appelquist:1980vg}
  T.~Appelquist and C.~W.~Bernard,
 {\it Strongly Interacting Higgs Bosons},
  Phys.\ Rev.\ D {\bf 22}, 200 (1980).

\bibitem{Longhitano:1980iz}
  A.~C.~Longhitano,
  Phys.\ Rev.\  D {\bf 22}, 1166 (1980).
  
\bibitem{Longhitano:1980tm}
  A.~C.~Longhitano,
  {\it Low-Energy Impact Of A Heavy Higgs Boson Sector},
  Nucl.\ Phys.\  B {\bf 188}, 118 (1981).

\bibitem{Appelquist:1993ka}
  T.~Appelquist and G.~H.~Wu,
 {\it The Electroweak chiral Lagrangian and new precision
  measurements},
  Phys.\ Rev.\  D {\bf 48}, 3235 (1993)
  [arXiv:hep-ph/9304240].

\bibitem{Chanowitz:1978uj}
  M.~S.~Chanowitz, M.~A.~Furman and I.~Hinchliffe,
  Phys.\ Lett.\  B {\bf 78}, 285 (1978).

\bibitem{Chanowitz:1978mv}
  M.~S.~Chanowitz, M.~A.~Furman and I.~Hinchliffe,
  Nucl.\ Phys.\  B {\bf 153}, 402 (1979).



\bibitem{Vayonakis:1976vz}
  C.~E.~Vayonakis,
  Lett.\ Nuovo Cim.\  {\bf 17}, 383 (1976).

\bibitem{Georgi:1985hf}
  H.~Georgi,
  Nucl.\ Phys.\  B {\bf 266}, 274 (1986).


\bibitem{Hill:2002me}
  C.~T.~Hill and A.~K.~Leibovich,
  Phys.\ Rev.\  D {\bf 66}, 016006 (2002)
  [arXiv:hep-ph/0205057].

\bibitem{Schwinn:2004xa}
  C.~Schwinn,
  Phys.\ Rev.\  D {\bf 69}, 116005 (2004)
  [arXiv:hep-ph/0402118].

\bibitem{Schwinn:2005qa}
  C.~Schwinn,
  Phys.\ Rev.\  D {\bf 71}, 113005 (2005)
  [arXiv:hep-ph/0504240].
  
\bibitem{Einhorn:1986za}
  M.~B.~Einhorn and G.~J.~Goldberg,
  Phys.\ Rev.\ Lett.\  {\bf 57}, 2115 (1986).

\end{thebibliography}
\end{document}